\documentclass[journal]{IEEEtran}
\usepackage{ifthen}
\usepackage{pifont}
\usepackage{amssymb}
\usepackage{acronym}

\usepackage[keeplastbox]{flushend}
\usepackage{xspace}
\usepackage[dvipsnames]{xcolor}
\usepackage{times}%
\usepackage[numbers]{natbib}
\usepackage{amsmath}
\usepackage{dcolumn}
\usepackage{endnotes}
\usepackage{graphics}
\usepackage{color}
\usepackage[small]{caption}
\usepackage{epsfig}
\usepackage{calc}
\usepackage{amssymb}
\usepackage{amstext}
\usepackage{amsmath}
\usepackage{amsthm}
\usepackage{multicol}
\usepackage{graphicx}
\usepackage{multirow}
\usepackage{pslatex}
\usepackage[caption=false]{subfig}
\usepackage{rotating}
\usepackage{lscape}
\usepackage{nameref}

\usepackage{amssymb}
\usepackage{cleveref}
\usepackage{booktabs}
\usepackage{eurosym}
\usepackage{inconsolata}
\usepackage{glossaries}
\usepackage{pgf,tikz}
\usetikzlibrary{arrows,automata,positioning,shapes.geometric,calc}

\newboolean{showtodos}
\setboolean{showtodos}{true} 
\ifthenelse{\boolean{showtodos}}
  {
    \newcommand{\todomarker}[4]{\marginpar{\textcolor{#1}{#2}}\footnote{
    \textcolor{#1}{\it\scriptsize {\textbf{\sf \underline{#3}:~#4}}}}}
  }
  {
    \newcommand{\todomarker}[4]{}
  }

\newboolean{showcomments}
\setboolean{showcomments}{true} 
\ifthenelse{\boolean{showcomments}}
  {
    \newcommand{\notemarker}[4]{{\textcolor{#1}{#2}}\footnote{
    \textcolor{#1}{\it\scriptsize {\textbf{\sf \underline{#3}:~#4}}}}}
  }
  {
    \newcommand{\notemarker}[4]{}
  }

\newcommand{\fra}[1]{\notemarker{violet}{\ding{74}}{francesco}{#1}}

\newcommand{\simu}[0]{Tangramob\xspace}
\newcommand{\simus}[0]{Tangramob\xspace}

\newcommand{\smarthub}[0]{\textit{tangrhub}\xspace}
\newcommand{\smarthubs}[0]{\textit{tangrhubs}\xspace}

\newcommand{\revisedtext}[1]{#1}

\ifCLASSINFOpdf
\else
\fi
\begin{document}
%
\title{Tangramob: an agent-based simulation framework for validating urban smart mobility solutions}
%
%
%

\author{\IEEEauthorblockN{Carlo Castagnari,
Flavio Corradini,
Francesco De Angelis,
Jacopo de Berardinis,
Giorgio Forcina,
Andrea Polini}\\
\IEEEauthorblockA{School of Science and Tecnology, University of Camerino, via Madonna delle Carceri 9, Camerino MC, Italy}
\thanks{Manuscript under revision.}%
\thanks{Corresponding author: F. De Angelis (francesco.deangelis@unicam.it).}}

%
%

\markboth{Internal version for arXiv.org}{}%

%



\maketitle

\begin{abstract}
   Estimating the effects of introducing a range of smart mobility solutions within an urban area is a crucial concern in urban planning. The lack of a Decision Support System (DSS) for the assessment of mobility initiatives, forces local public authorities and mobility service providers to base their decisions on guidelines derived from common heuristics and best practices.
   These approaches can help planners in shaping mobility solutions, but given the high number of variables to consider the effects are not guaranteed. Therefore, a solution conceived respecting the available guidelines can result in a failure in a different context. In particular, difficult aspects to consider are the interactions between different mobility services available in a given urban area, and the acceptance of a given mobility initiative by the inhabitants of the area.
   
   In order to fill this gap, we introduce \simu, an agent-based simulation framework capable of assessing the impacts of a Smart Mobility Initiative (SMI) within an urban area of interest. \simu simulates how urban traffic is expected to evolve as citizens start experiencing the newly offered traveling solutions. This allows decision makers to evaluate the efficacy of their initiatives taking into account the current urban system.
   In this paper we provide an overview of the simulation framework along with its design. To show the potential of \simu, 3 mobility initiatives are simulated and compared on the same scenario. This shows how it is possible to perform comparative experiments so as to align mobility initiatives to the user goals.
\end{abstract}


%
\IEEEpeerreviewmaketitle


\section{Introduction\label{cha:introduction}}

According to the United Nations \cite{nations2016world}, in 2016, world's population was 7.4 billions inhabitants and about 54.5\% of them lived in urban areas. 
Despite all the benefits historically brought by urbanization, like poverty reduction, longer life expectancy and economic wealth, such an uncontrolled demographic growth is pushing cities to deal with several management problems.
In particular, focusing on urban mobility, transport infrastructures are close to saturation and this comes with a bunch of problems like car dependence, spatial footprint, traffic congestion, air and noise pollution. Novel smart mobility solutions need to be introduced, and investments have to be carefully assessed in relation to their effective potential to improve the mobility ecosystem.


Novel mobility initiatives are generally shaped, and their adoption assessed, considering common guidelines and best practices. Nevertheless it is not seldom the case that the observed effects, after the concrete deployment of a solution, are not satisfactory. In particular, there are two complex aspects that are difficult to assess when following such approaches to planning. The first one relates to how the new mobility solution will interact with the already available ones, whereas the second one relates to citizens acceptance.
Indeed as many articles report (\cite{pitas2014, alter2013, westneat2016, negri2016, mamiit2016, sulopuisto2016, fehrenbacher2013}), there are many cases in which the adoption of a smart mobility initiative did not bring the expected benefits. In particular, many real scenarios from Europe (\cite{negri2016, sulopuisto2016, pitas2014}), China (\cite{mamiit2016}) and U.S.A (\cite{alter2013, westneat2016, fehrenbacher2013}) demonstrated how proposed smart mobility services failed, since they were not accepted by communities.
As argued by \cite{pitas2014}, the carsharing failure in London is attributed to a bad service configuration. Similarly, \cite{alter2013, westneat2016, negri2016} argue how initially promising bikesharing solutions for the city of Salerno (IT) and Seattle (USA) have not been adopted by the population. The lack of a formal way to estimate the impacts of a range of smart mobility services and their interactions is also reported in \cite{McGrath2016} and \cite{Zavitsas2011}, remarking both the importance, and the actual absence of a ``common framework'' for this purpose.
Indeed, decision makers are in urgent need of innovative approaches providing quantitative forecasts in relation to the different aspects connected with the introduction of a novel mobility initiative. Resulting Decision Support Systems (DSS) will complement already available approaches in the definition and shaping of the smart mobility solution to adopt.





This considerations motivated us in developing a novel DSS named \simu. This is an agent-based simulation framework capable of assessing the impacts of the introduction of a novel smart mobility initiative (i.e. a range of either homogeneous or heterogeneous smart mobility services) within an urban area of interest taking into account the current mobility ecosystem, as well as 
salient features of citizens in relation to the usage of mobility services.
Indeed, agent-based approaches are considered effective for searching a solution within huge state spaces when the domain to represent can be easily conceived as a composition of heterogeneous entities interacting in a distributed setting \cite{davidsson2007,barbati2012}. This is certainly the case of a mobility ecosystem in which many different entities can be identified, each one with its specific characteristics (e.g. commuters, transport means, roads, etc.), and the system behaviour and its features emerge from the interactions among such entities. 
\simu will simulate one day of mobility starting from a description of the population of interest and of the mobility resources available in the considered area, including the ones related to the initiative to evaluate. The day will be simulated many times over many iterations to derive a final configuration and the corresponding output. The iterations are needed since learning mechanisms are applied in \simu to let commuter agents try out the different available mobility solutions. After each iteration each commuter will provide a score for the travel experience according to its own profile, and taking into account quantitative parameters, such as travel times. In this way, a commuter agent will learn which are the transportation solutions that better fit its profile. Notably, the possibility to change transportation means in relation to the different segments of a travel allows intermodality  and multimodality within a simulation.
Clearly, the more the provided data input are effectively representative of the reality of interest, the more the returned data set will approximate the possible effects introduced by the mobility solution under evaluation. From the output data it will be possible to derive quantitative analysis in relation to changes in emissions, costs etc, as detailed in the following sections.
Summarizing, \simu is a DSS that helps decision makers in planning SMIs. The DSS is distinguishable from other proposals in relation to two main aspects: (i) it supports the simulation of intermodal and multimodal transport services; and (ii) 
it makes it possible to reflect the diversities of commuters with respect to their personal characteristics (e.g. gender, age, travel demand).



The simulator is 
built over MATSim, a powerful traffic simulator \cite{matsimbook}. \simu is aimed at all people involved in defining and planning new mobility services: urban planners, who are in charge of improving urban mobility; transport companies, which need to ponder their investments; researchers, aiming at testing and validating new mobility solutions. 

The rest of the paper is organized as follows: Section \ref{cha:idea} outlines the idea behind the \simu simulator and how it is expected to address the research problem. In section \ref{sec:smarthubABM} we provide an overview of the agent-based model of \simu and section \ref{cha:design} describes its architecture. Finally, section \ref{cha:results} proves the effectiveness and the potentialities of \simu by reporting an example of use. Section \ref{cha:related} shows the current attempts in supporting urban planners and mobility service providers in urban mobility planning. and section \ref{cha:conclusions} reports some conclusions and opportunities for future work.

\section{The \simu simulator}
\label{cha:idea}

\simus is an agent-based simulation framework that intends to support public and private decision makers in the task of shaping smart mobility initiatives for a specific urban area of interest. 
%
It can be considered as a Decision Support System (DSS) for smart mobility validation, focusing on the ability to capture and reproduce the mobility behaviour of each single commuter belonging to the selected sample population. For this purpose, \simus is organized as a simulation environment that the urban planner can easily use in order to understand if introducing a smart mobility initiative, i.e. a collection of mobility services, can improve the traveling experience of citizens as well as the performance of the urban transport system. 
Since the simulator is based on an Agent-Based Model (ABM), for each person in the sample population, represented as an autonomous reasoning agent, we can observe whether or not it will make use of the new mobility services. These fine-grained results also provide users with a measure concerning the expected adoption rate of the simulated mobility initiative, so as to figure out beforehand if the initiative can potentially succeed or not.

Technically, a \simus simulation requires four inputs: 
\begin{itemize}
    \item the \textbf{urban road network} of the area under study, 
    \item a representative population of the area with the \textbf{mobility agendas} of people. An agenda summarizes what a person does during an ordinary working day (i.e. activities) and how he moves from one place to the next one (i.e. \textit{legs}).
    \item the description of the \textbf{mobility services already offered} by the city: public transport timetable, etc.;
    \item the \textbf{smart mobility initiative} to evaluate, that is a list of geographically located containers (called \smarthubs) of one or more smart mobility services. Each smart mobility service belongs to a \smarthub and it comes with a number of mobility resources (e.g. vehicles), as well as a service charge (i.e. cost per km and cost per hour).
\end{itemize}
%
%
It is worth mentioning that the definition of an agent population is certainly the most complex and critical information to supply, in particular in relation to profiles, and details on daily travels. Obviously, the more the population is representative of the reality of interest the more the results of the simulation can be considered a good approximation. Strategies for the derivation of a population are out of scope for this paper, nevertheless different sources are available to define a representative synthetic population. Relevant data can be certainly collected from periodic census or questionnaires distributed to a sample population. Particularly effective nowadays is Mobile Crowd Sensing (MCS) \cite{Guo2015} that uses mobile apps developed for large scale sensing, and involve the contribution from a crowd of people that behave as \textit{probes} of the dynamics of the mobility in the city \cite{ma2014opportunities}. GPS data produced by the crowd are an excellent source of planning and transport information, and they are widely used in mobility project (e.g., the community based GPS navigator Waze \textit{(www.waze.com)} that tracks users to understand roadway congestion).
Activity recognition of travel demand models can also be derived 
using Input-Output Hidden Markov Models (IOHMMs) to infer travelers' activity patterns from call detail records as suggested in \cite{yin2017generative}.
%
%

Starting from the provided information, the execution of \simus can be thought of as performing a comparative experiment. The experiment consists in introducing the smart mobility initiative (i.e. applying the treatment) into the urban area of interest (i.e. the treated system) while observing the same reality as it is today, namely with no smart mobility initiative (i.e. control system). In the end, we can observe how these systems differ with respect to the following measures:
\begin{itemize} 
    \item \emph{travel distance}, expressed in metres and referred to the distance traveled by each commuter;
    \item \emph{travel time}, expressed in seconds and referred to the time spent traveling for each commuter;
    \item \emph{$CO_2$ emissions}, expressed in grams and referred to the quantity of $CO_2$ produced by each commuter according to the used means of transport;
    \item \emph{cost of mobility}, expressed in euros and referred to the cost of mobility for each commuter;    
    \item \emph{urban traffic levels}, expressed as the number of traveling vehicles on each road at a given moment in time band. This statistic is a picture of the road infrastructure under study and it is useful when one needs to understand which are the most congested roads within a time slot. 
\end{itemize}

Such a comparison would allow the user to understand if the proposed mobility initiative is in line with their expectations. In case they are not satisfied with the achieved results, the user can change the configuration of the mobility initiative (e.g. relocating \smarthubs, adding/removing one or more \smarthubs, modifying the parameters of a mobility service and so forth) in order to repeat the experiment as before.



\subsection{The tangrhub} \label{ssec:smarthub}

In \simu, the actual placement of smart mobility services within the urban area under study is made possible by \smarthubs. A \smarthub can be defined as a geo-located entity providing citizens with one or more mobility services. 
\revisedtext{A \smarthub collects} one or more smart mobility services, each of which is offered by either private or public providers. For instance, a carsharing service provided by two different companies, results in two different characterizations of resources and their usage deployed within the \smarthubs of interest. 
Considering the typical urban conformation, such a flexible and modular abstraction allows urban planners to represent all the existing transport facilities like railway stations, bus stops and so forth, and to introduce intermodality among the  mobility services. Indeed, a bus stop could be represented as a \smarthub where only the bus service is available.


Examples of smart mobility services that the user can add to a \smarthub are: dynamic public transport, shared transport services (e.g. carsharing, bikesharing), dynamic ridesharing, autonomous taxis and so forth \cite{sharingmobit2016}. However, each smart mobility service $m_i$ provided by a \smarthub $th_j$ must belong to only one of the following service types:
\begin{itemize}
    \item \textbf{intra-hub} services, used for moving people \textit{to} and \textit{from} $th_j$ thereby serving first mile trips, e.g. from a commuter's home-place (departure) to $th_j$, as well as last mile trips, e.g. from $th_j$ to a commuter's workplace (destination).
    \item \textbf{inter-hub} services, for moving commuters from $th_j$ to another \smarthub $th_k$ supporting the service type of $m_i$.
\end{itemize}

From the simulator's perspective, we can think of a \smarthub as an entity with which people interact every time they need to travel. As a result of such interactions, \smarthubs are expected to collaborate with each other in order to provide commuters with a list of valid traveling solutions. Thus, it is up to each person to evaluate and choose the most suitable solution according to their own needs and preferences.

\subsection{Smart Mobility Initiative (SMI)} \label{ssec:smi}

According to the concept of \smarthub seen before, shaping a Smart Mobility Initiative (SMI) is about placing a number of \smarthubs within the urban area of interest, adding one or more smart mobility services to each of them, and providing a specific characterization for the added mobility services. 
Thus, it is up to the user (e.g. an urban planner) to design a list of candidate SMIs according to the goals and the available financial resources of his local authority.


To define a smart mobility service for a \smarthub, such as carsharing, the user has to specify the service type (i.e. intra-hub or inter-hub), the initial number of vehicles and the service charge (i.e. cost per km, per hour, and fixed), the $CO_2$ footprint, and other parameters depending of the type of vehicles.
Therefore, in \simu, a mobility service provided by an organization is represented as a whole as the sum of all the services made available by the same organization within the selected \smarthubs.
It is worth noticing that the cost of a mobility service does not need to correspond to a real currency. In fact, we can consider cost in terms of ``points" since such an approach fits the idea of \emph{mobility as a service} \cite{Finger2015, CatapultTransportSystems2016}.

These cost-related parameters are expected to affect the mobility decisions of commuters. More precisely, commuters are more inclined to choose the most convenient services, i.e. the ones with the greatest \textit{efficiency}/\textit{cost} trade-off. Thus, leveraging the cost of mobility services allows urban planners to achieve a mobility policy, thereby promoting some services against others. For instance, a cheap bikesharing service would hopefully be more preferable for commuters than an expensive carsharing service in case of short trips.

\revisedtext{\subsection{\simus commuting patterns} \label{ssec:smarthubcompat}}
 
\begin{figure}[htbp!]
  \centering
  \begin{minipage}{.49\linewidth}
  \centering
  \includegraphics[width = 0.95\linewidth]{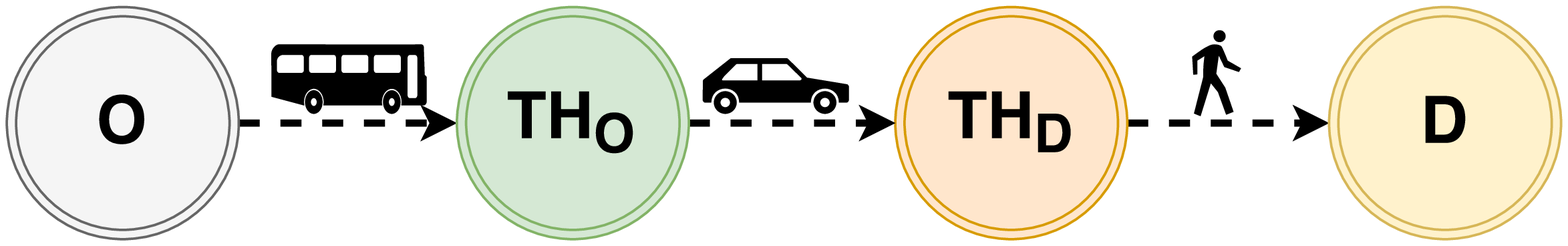}
  \caption{3-trip path\vspace{1em}}  
  \label{fig:originHubHubDestinationPattern}
  \end{minipage}
  \begin{minipage}{.49\linewidth}
  \centering
  \includegraphics[width = 0.95\linewidth]{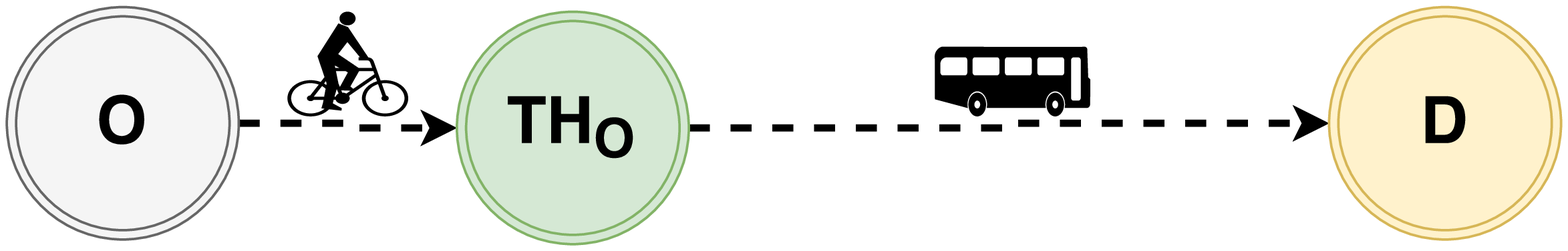}
  \caption{2-trip path I\vspace{1em}}
  \label{fig:originHubODestinationPattern}
  \end{minipage}
  \begin{minipage}{.49\linewidth}
  \centering
  \includegraphics[width = 0.95\linewidth]{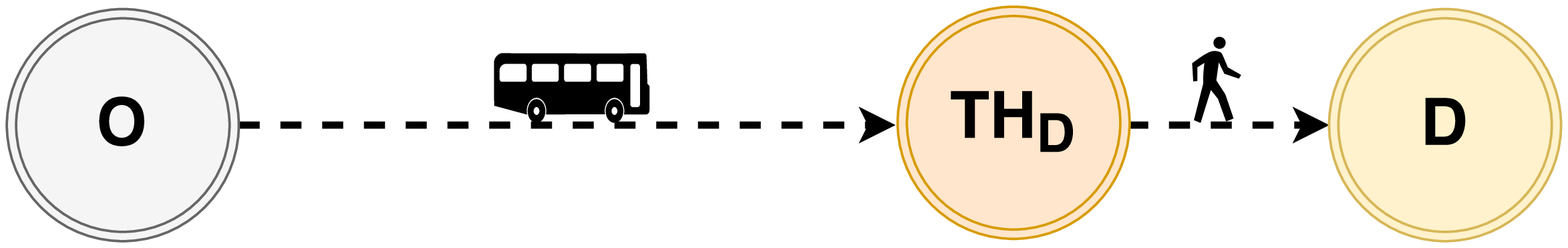}
  \caption{2-trip path II}
  \label{fig:originHubDPattern}
  \end{minipage}
  \begin{minipage}{.49\linewidth}
  \centering  
  \includegraphics[width = 0.95\linewidth]{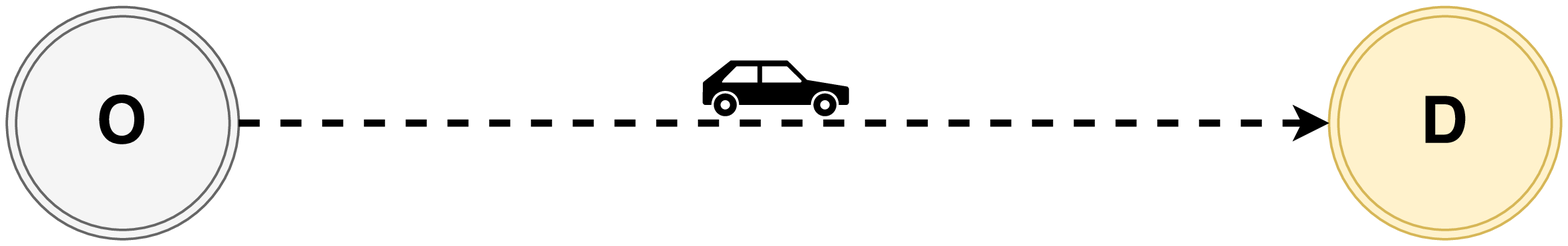}
  \caption{Direct path}
  \label{fig:originDestinationPattern}  
  \end{minipage}
\end{figure}

As new mobility opportunities are introduced, commuters are expected to change their daily commuting patterns.
A \textbf{commuting pattern} is the intermodal representation of how a person moves from one place to another. Such a trip can be either simple (e.g. by car) or more complex (e.g. by a combination of travel modes). A clear example of a commuting pattern is a route provided by Google Maps.

In \simu, the complexity of commuting patterns can be limited thanks to the direct interconnection of \smarthubs via their \emph{inter-hub} mobility services. 
Commuters are never offered with traveling solutions made up of more than three sub-trips.

Letting $TH_o$, $TH_d$ be tangrhubs respectively close to origin $o$ and destination $d$ of a trip, we can group all the possible commuting patterns in four classes. In the first class (Figure \ref{fig:originHubHubDestinationPattern}) a person $p$ is expected to reach $TH_o$ either by walk or using an \emph{intra-hub} mobility service provided by the same \smarthub; once arrived at $TH_o$, an \emph{inter-hub} mobility service will bring $p$ at $TH_d$; finally, the commuter will be able to reach his destination $d$ either by walk or using an \emph{intra-hub} mobility service offered by $TH_d$.
Analogously, the second and the third classes (Figures \ref{fig:originHubODestinationPattern} and \ref{fig:originHubDPattern}) represent a combination of two modal trips performed either by \emph{intra-hub} services or by personal traveling modes.
Finally, the last class (Figures \ref{fig:originDestinationPattern}) corresponds either to a direct trip (e.g. by car, bike, walk), or to the case a single \emph{inter-hub} service is used.


\section{\simu Agent-Based Model overview} 
\label{sec:smarthubABM}

Starting from the idea of \simu, we  present the Agent-Based Model (ABM) on which it is conceived. 
The \simu ABM is composed of two agent types: \emph{commuter} and \emph{\smarthub}. A \emph{commuter} agent is the computational representation of a single person that is part of the sample population under study.
Every commuter agent comes with some relevant personal characteristics, like gender and age, affecting the outcomes of the actions taken during the simulation. These effects also impact on the behavior of commuters. For instance, an elderly person will be less prone to travel by bicycle for long trips, since this would take too long for him.
More importantly, every commuter has a personal mobility agenda, i.e. a sequence of daily activities (e.g. home, work, etc.) interleaved by mobility segments that tells how the agent manages to get from one activity location to the next one. 

On the other hand, a \smarthub agent can be defined as a local mobility service provider with the ability to improve its services as the simulation iterates; in the real-world, this active behaviour might corresponds to a daily enhancement.

Both agents live and operate, albeit with different perceptions, in a composite environment that is made of three different spaces: the temporal space, the geographical space and the smart mobility services' state space. Specifically, the temporal space 
reflects the passage of time in seconds. The geographical space can be defined as the directed weighted graph resulting from the road network infrastructure of the urban area under study; in particular, nodes represent intersections and edges denote streets. 
Such a space is the actual core of the transport simulation, since the physical limitations of the road infrastructure can create bottlenecks and delays as people move with a certain pace. Finally, the last sub-environment is meant to represent the status of all the smart mobility services which are currently provided by \smarthubs. This space can be conceived as a \emph{tuple space} in which the status of each mobility service breaks down into a number of smaller sub-states. For instance, the status of a carsharing service can be expressed as the combination of the states of all its vehicles.

\begin{figure}[ht]
    \centering
    \includegraphics[scale = 0.4]{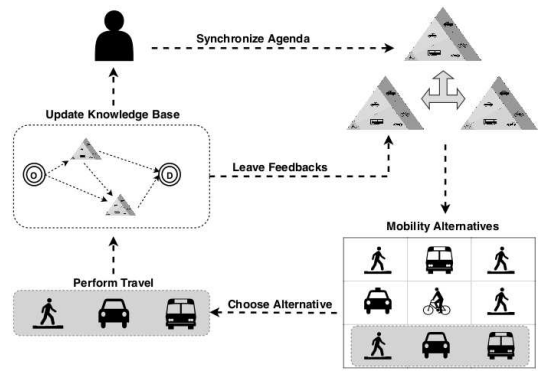}
    \caption{Commuter-Tangrhub interaction loop}
    \label{fig:commuter-tangrhub interaction}
\end{figure}

This complex environment allows agents to perform actions that can eventually alter the state of affairs of one or more sub-environments. In particular, as depicted in Figure \ref{fig:commuter-tangrhub interaction}, every time a commuter needs to move from one place to another, an interaction with the surrounding \smarthubs takes place. During this interaction, a smart mobility negotiation occurs: the \smarthubs collaborate with each other in order to provide the commuter with a number of traveling alternatives. A traveling alternative can be thought of as a combination of one or more (up to three) mobility segments, each of which can involve a smart mobility service and it is based on the \simu commuting patterns seen in section \ref{ssec:smarthubcompat}. Next, the commuter agent will perform a decision-making process so as to select the traveling alternative that is expected to maximize his performance criteria. 

The alternative selection process is organized  as  follows:  first,  every  single  travelling  alternative is evaluated according to the expected performance of each segment  it  is  made  of;  
then, the cost is introduced to influence such preference-ordered rank; finally, a travelling alternative is selected and then simulated. Once  the commuter agent has reached his final destination, he  is  expected  to assign  a  score to every  single  commuted  mobility segment to record its traveling experience so as  to  make  more  informed  decisions  for the next iterations. 

As soon as a traveling alternative is chosen, the involved \smarthubs will reserve the required mobility services so that the commuter can start his journey. Finally, once the commuter has reached his destination, he will be asked to leave a feedback for each smart mobility service used.  


The behaviour of a commuter revolves around four actions: (i) synchronizing his mobility agenda with the closest tangrhubs, to obtain a list of traveling alternatives for reaching the location of the forthcoming activity; (ii) choosing a traveling alternative out of the proposed ones; (iii) performing the chosen traveling alternative; and finally (iv) leaving as many feedback as the number of mobility services used in the course of the day. A commuter will then try to maximize his/her traveling experience minimizing travel time, covered distance, emissions and the cost of mobility. More precisely, this is done by selecting the traveling alternative that is expected to optimize such criteria from time to time.

On the other hand, the \smarthub agent has the following two goals: to maximize the traveling experience of commuters and minimize the number of mobility resources for each service. Thus, in order to achieve these objectives, a \smarthub can perform the following actions: (i) building a list of traveling alternatives in collaboration with other \smarthubs, (ii) provide a commuter with a list of valid traveling alternatives, (iii) update the status of its own mobility services, (iv) improve and optimize its own mobility services.

The \smarthub's service adaptation process is made possible by commuters feedback. In particular, each feedback qualifies the traveling experience of a commuter using a specific mobility service. 
Collecting and averaging all the feedback of a mobility service can give a metric concerning the performance of that service, thereby contributing to its improvement and optimization. For instance, if all the daily-collected carsharing feedback are negative, a \smarthub would have a valid indicator of such an inefficiency to run for cover.
Therefore, the purpose of a feedback is twofold: on one hand, it pushes the commuter agent to reason about the quality of the mobility services to make more informed decisions for the next iterations; on the other hand, it enables \smarthubs to align to the actual mobility needs of the population.


\simu simulations are thus iterative: each iteration corresponds to a typical day in which commuters experiment with the introduced smart mobility services in order to record their performance, while \smarthubs can improve their services iteration by iteration. This time-evolving behaviour, driven by feedback, enables commuters to make more informed decisions every time they are offered a list of traveling alternatives. 
Therefore, commuters are modeled as \emph{proactive agents} since there is need of an iteration-persistent memory structure, i.e. a \textit{knowledge base}, to implement such an experience-based learning capability. With that idea, the decision-making process of commuters exploits their personal knowledge base in order to evaluate the expected score of a traveling alternative, thanks to the experience accumulated from past iterations. 
This is achieved by updating the knowledge base, by means of a Hebbian-like learning function. This will permit to gradually accumulate scores so as to let the commuter maturate an experience-based perception for every segment.
Similarly, \smarthubs are modeled as \textit{self-adaptive agents} which can use different strategies and optimization methodologies 
to enable their travel improving behavior
at the end of each iteration and by means of feedbacks. 




\section{Design \& Implementation Overview}
\label{cha:design}

Considering the \simus agent-based nature, the framework has been developed on an already validated and robust agent-based traffic simulator named MATSim  \cite{matsimbook} (Multi Agent Transport Simulation). Such a design choice is due to the fact that it is possible to represent the characterizations and the behaviors of both our agent types in MATSim. Moreover, such a simulator can be adapted to support all the sub-environments of the model, allowing \simus to evaluate the performance criteria as outcomes from the interactions among such spaces 
and agents. 

\subsection{Multi Agent Transport Simulation: MATSim}

\begin{figure}[!htbp]
    \centering
    \includegraphics[scale = .2]{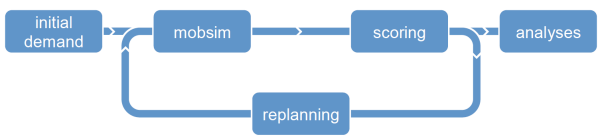}
    \caption{MATSim modules}
    \label{matsimModule}
\end{figure}

MATSim \cite{matsimbook} is an activity-based multi-agent simulation framework for implementing large-scale agent-based transport scenarios. It is an open-source project implemented in Java under the GNU public license. As in Figure \ref{matsimModule}, the framework consists of several modules which can be combined, used standalone, or replaced by own implementations.

MATSim is designed to model a single day and it is based on a co-evolutionary approach in order to reproduce real-life scenarios. Every agent repeatedly optimizes its daily activity schedule while in competition for space-time slots with all other agents on the transportation infrastructure. This optimization follows an iterative process and it is based on different choice dimensions such as route selection, time choice and mode choice. A MATSim run consists of a number of iterations in which the steps in Figure \ref{matsimModule} are repeated in a cyclical manner.
The \textit{initial demand} arises from the study of the area to simulate, and it comprehends its topology (i.e. the network), the mobility habits of its inhabitants and their personal features. Every citizen possesses a memory containing a fixed number of daily plans, each of which is composed of a daily activity chain and an associated score, i.e. the utility of that plan. Once the features of the scenario's components are acquired, and before the MATSim mobility simulation, every agent selects a plan from its memory according to the score associated with each of them: higher score plans are more likely to be selected. 

Afterwards, the selected plans are executed by means of the mobility simulation module; the latter relies on the concept of queue simulation, which was demonstrated to efficiently approximate real-life traffic flows. In particular, MATSim models roads as FIFO queues with a limited vehicle capacity; every time a vehicle asks permission to access a road, the corresponding road agent can either respond positively, in case there still is space, or negatively, if the queue capacity is reached. Thus, in case a commuter is not allowed to enter a road segment a delay is produced in the system as long as the regular flow is restored. Finally, when a commuter manages to enter a road, he is added to the queue tail until he reaches its head in order to move to the next segment. 

The score computation is made after every \textit{mobsim} run, and it is performed on the last executed plans by means of a \textit{scoring} function. The score represents a measure about how the traveling choices made by agents affected the execution of their activities: the higher the score the better the day.

Once all plans have been scored, the \textit{replanning} phase takes place, as a portion of the population is allowed to modify their plans, in accordance with the co-evolutionary approach. Then, such plans are modified by applying a mutation operator according to the previously mentioned choice dimensions.

Finally, in the last iteration the \textit{replanning} phase is not executed anymore and it is replaced with the \textit{analysis} module. In fact, MATSim is strongly based on events stemming from the \textit{mobsim} and this allows to record every action in the simulation for further \textit{analysis}. These events' records can be aggregated to evaluate any measure at the desired resolution.

MATSim can be applied in large scenarios. We show an example considering a small city in the paper, nevertheless scalability to bigger cities should not be much a problem since MATsim simulations of large-scale agent-based micro simulation models are proved scalable \cite{Waraich2015}. 
%
%
An experiment made by MATSim developers with 1.62 million agents and 163K links in the area of Zurich city were simulated in about 20 minutes in a machine with 128GB RAM and 8 dual-core AMD Opteron CPUs. Also the Switzerland traffic was modeled in about 3 hours for a single MATSim iteration: one million roads and 7.3 million agents clearly show that large-scale, multi-agent micro-simulation can reasonably be used.

\subsection{\simu meets MATSim}

Our framework has been implemented on top of MATSim, taking advantage of its flexible and modular architecture and trying to maintain the same design principles. We redefined and extended the behavior of some original MATSim modules, whereas other remarkable contributions were introduced in such a way to capture all the features of the \simu AB model of Section \ref{sec:smarthubABM}.

\begin{figure}[ht]
    \centering
    \includegraphics[width=.78\linewidth]{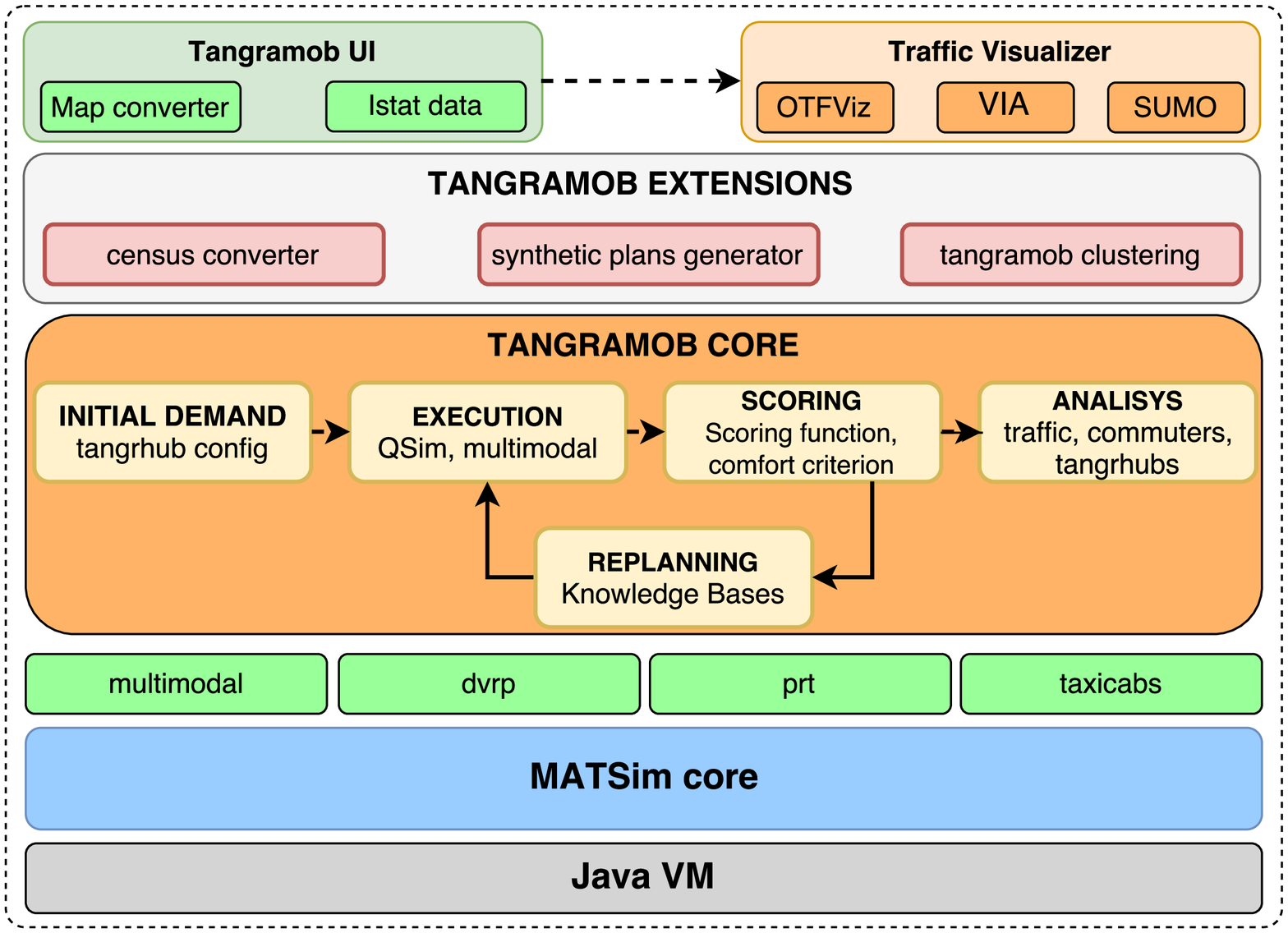}
    \caption{\simu architecture}
    \label{architecture}
\end{figure}

In particular, the module \emph{initial demand}, in which the simulation input data are collected and validated, is integrated with the specification of the SMI, describing the locations of the \smarthubs on the map as well as the list of the mobility services available on each of them. Making this integration possible required us to implement the concepts of \emph{\smarthub agent}, a new static but active entity that is responsible both for managing and offering new traveling opportunities to the nearby population, and for managing the associated mobility services, which can be seen as services provided by private or public companies/organizations and that overall constitute the infrastructure of the SMIs available in the urban area. 

The \emph{mobsim} module, specialized in simulating the urban traffic, has been integrated with the MATSim ``multimodal" extension, that allows to deal with different transport modes as well as simulating the overtaking of vehicles. This way, \simu can also evaluate the impact on the urban system caused by unconventional kinds of vehicles (e.g. scooters, bicycles, etc.). For this purpose, we redesigned the original concept of MATSim's vehicles, and we introduced the characterization of mobility services with the ability to manage such vehicles.
Furthermore, our characterization takes into account the most relevant vehicles features like: dimension, velocity, fuel type and consumption; all these specifications are expected to impact on the traffic simulation, especially for what concerns travel delays and times, and thus are relevant information in relation to the mobility decisions of commuters.


Concerning the \emph{scoring} module, \simus still exploits the original Charypar-Nagel scoring function \cite{matsimbook}. This allowed us to validate the new learning process of Tangramob, exploiting the existing MATSim's validation work.


The \emph{replanning} phase designed for \simus is completely different from that followed by MATSim. Whereas MATSim adopts a co-evolutionary algorithm, our framework is based on a reinforcement learning approach, allowing each commuter to evaluate his past traveling experience in order to improve their daily personal mobility. This is made possible by the implementation of iteration-persistent memory structures, that every commuter can exploit as knowledge base, in order to accumulate the score given for each mobility service used during the simulation. Thus, the score of a service acts as a reward for the action of choosing that service for a certain trip. Such a different approach allows commuters to maximize the expected utility of their mobility decisions. In particular, during the last iteration of the simulation, each commuter will decide to either use the new mobility services, or not to accept the mobility initiative, according to the collected knowledge.


Finally, the \emph{analysis} module has been integrated with new statistical collectors to gather all data useful to compare the legacy urban mobility with the one emerging after the introduction of a SMI. Some stats correspond to the agents' performance criteria described in section \ref{sec:smarthubABM}, and others are focused on the urban system as a whole. In particular, we aim at collecting the following statistical data: (i) urban traffic levels, (ii) ${CO}_2$ emissions, (iii) traveled distances, (iv) travel times, (v) land use levels, (vi) cost of mobility, (vii) number of adopters, and (viii) resource usage level. 


As depicted in Figure \ref{architecture}, all these redefined modules form the core of \simu, sitting on the top of MATSim and some of its well-known extensions. Moreover, to make the simulator more accessible and user-friendly, we have designed and implemented the following features:
\begin{itemize}
    \item a \emph{census data converter}, namely a  tool for translating Italian census data into suitable input mobility agendas;
    \item a \emph{population generator}, for synthesizing a sample population from some statistical data about an urban area for which neither census nor plans are available;
    \item \emph{\smarthub aided placement}, a tool that analyzes the locations of people daily activities in order to spot the most populated urban areas. By using clustering algorithms, this tool supports users in the task of placing \smarthubs in a more rational way, since keeping them close to the population is expected to minimize micro-mobility.
    \item a web-based \emph{Graphical User Interface (GUI)}, which allows users to select the geographical area of interest 
    to automatically retrieve the urban road network of the selected area from \textit{OpenStreetMap}.
    In case the user cannot generate the data of the sample population under study from the census data converter, the user is expected to load the mobility agendas manually or to generate a synthetic but realistic population from other sources.
    Finally, once the geographical context is defined, the user can shape a smart mobility initiative by geographically placing a number of \smarthubs and configuring each of them with one or more mobility services.
\end{itemize}


The resulting architecture is fully extensible in every layer, providing the possibility to develop extensions over both the MATSim layer and the \simus codebase.


\section{\simu: an example of use}
\label{cha:results}
In order to show an example of use of \simu, we report some experiments performed on a real scenario in the city of Ascoli Piceno (Italy).
Ascoli Piceno is a small city with about 50.000 inhabitants over 158 $km^2$ and several other thousand of people who live in near places outside the city perimeter.


\begin{figure}[ht]
    \centering
    \includegraphics[width=.965\columnwidth]{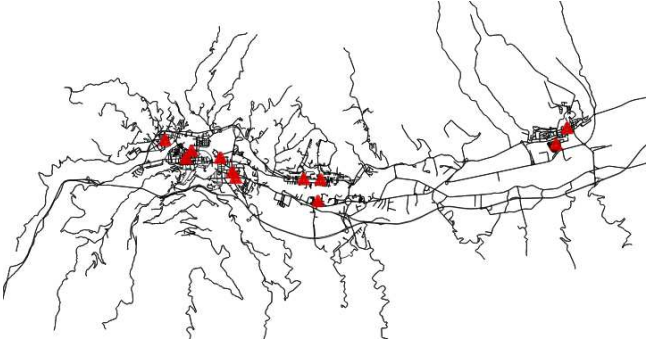}
    \caption{Ascoli Piceno network with \textit{tangrubs} positioning for SMI's}
    \label{gridNet}
\end{figure}


As depicted in Figure \ref{gridNet} the network represents all the city roads and infrastructures including the city center and the roads that connect the city with other places. Starting from available statistics, we identified 15 areas  and for each one we identified inhabitants and jobs as described in Table \ref{tab:areas}.
%
\begin{table}[ht]
    \centering
    \begin{minipage}[t]{0.485\columnwidth}
    \begin{tabular}{l|c|c||}
     \multicolumn{1}{c|}{ Areas} & People  &   Jobs\\
     \hline
     P.ta Solest\'a   &   5009    &      3170\\
     P.ta Romana       &   1839    &       700\\
     Centro      &   7740    &      6760\\
     Piazzarola         &    409    &       250\\
     C. Parignano    &   3368    &      2170\\
     P.ta Maggiore      & 11500     &      10900\\
     Monticelli         &  10633    &      8000\\
     Brecciarolo        &    645    &      1300\\
    \end{tabular}
    \end{minipage}
    \begin{minipage}[t]{0.485\columnwidth}
    \begin{tabular}{l|c|c}
     \multicolumn{1}{c|}{ Areas} & People  &   Jobs\\
     \hline
     P. di Bretta      &   1694    &       500\\
     Battente        &    103    &      3000\\
     Marino     &    576    &      1400\\
     Villa Pigna   &   3000    &      2000\\
     Z. Industriale   &    500    &      6500\\
     C. di Lama    &   3000    &      5000\\
     Frazioni           &   6242    &      6000\\
     \hfill & \hfill & \\
    \end{tabular}
    \end{minipage}
    \caption{Population and jobs in the city areas}
    \label{tab:areas}
\end{table}

In the experiment, we modeled the whole population considering the suburbs with 56.000 agents. Using the statistics of the municipality we have built a normally distributed population age with a 45\% in range 25-49. Female are 52\% and male 48\%.
These parameters are expected to affect the act of travelling of commuters, thus impacting on their score.
Basically,  mobility agendas has been organized with three daily activities in the following order: \textit{home, work, home}. Thus, a commuter moves from its home to a workplace in another area and viceversa. For sake of clarity, we consider in this experiment as \textit{work} each kind of activity different to stay at \textit{home}. We also do not consider multi-trip commutes.


As a typical real-case scenario, peak activity hours can be split into two different moments: 8:00 a.m. commuters start moving towards the workplaces; while at 16:00 p.m. commuters come back home from work. The first activity in the morning is distributed in the 5:00 a.m. - 13:00 a.m. time slot with 45\% included in the 07:00 a.m. - 08:00 a.m. hour. The homecoming happens at the end of the \textit{work} activity. That time is modeled using a Gaussian distribution centered over 6 hour of duration.

In this scenario, we aim at investigating the impacts of three different smart mobility initiatives that integrate  transport services: a bikeshare, an electric carshare, and an e-scootersharing service. All vehicles used are zero emissions. We use 11 \smarthubs in the city areas  as several locations in the city center can be served by the same hub. Each hub is characterized as in Table \ref{tab:grid}, for readability we use the same resources for each \smarthub in this example.

As shown in Table \ref{tab:grid}, each SMI shares the same number of \smarthubs, each of which is provided with the same choice set of mobility services. Even the geographical location of \smarthubs is the same for all the initiatives, and it is denoted by the triangles depicted in Figure \ref{gridNet}. For each \smarthub we specify the dimension of the fleet the hub manage at the start of the simulation, its total capacity to store vehicles is set a 25\% more than the initial fleet. What differs among these initiatives is just the number of mobility resources, which in this case, correspond to the vehicle fleet of each service.

\begin{table}[ht]
    \centering
    \begin{tabular}{cc|c|c|c|}
        \cline{3-5}
        \multicolumn{1}{l}{}                        & \multicolumn{1}{l|}{} & \multicolumn{3}{c|}{Fleet} \\ \hline
        \multicolumn{1}{|c|}{}             & Mobility Services     & SMI-1   & SMI-2   & SMI-3  \\ \hline
        \multicolumn{1}{|c|}{\multirow{3}{*}{TH}} & bikesharing        & 10       & 10       & 50      \\ \cline{2-5} 
        \multicolumn{1}{|c|}{}                      & carsharing            & --       & 10       & 50      \\ \cline{2-5} 
        \multicolumn{1}{|c|}{}                      & e-scootersharing      & --       & 10       & 50      \\ \hline
        \multicolumn{1}{|c|}{\multirow{3}{*}{Total}} & bikesharing        & 110       & 110       & 550      \\ \cline{2-5} 
        \multicolumn{1}{|c|}{}                      & carsharing            & --       & 110       & 550      \\ \cline{2-5} 
        \multicolumn{1}{|c|}{}                      & e-scootersharing      & --       & 110      & 550      \\ \hline
    \end{tabular}
    \caption{Grid network: \smarthubs experimental setup}
    \label{tab:grid}
\end{table}

For each mobility service we specify costs. For this experiment, we set the costs of the chosen mobility services, which in turn were set according to the actual average service charges in Europe, as summarized in Table \ref{tab:costs}.
\begin{table}[ht]
\centering
\begin{tabular}{|l|c|c|c|}
   \cline{2-4}
 \multicolumn{1}{l|}{}     & \multicolumn{1}{l|}{Cost per h} & \multicolumn{1}{l|}{Cost per km} & \multicolumn{1}{l|}{Fixed Cost} \\ \hline
Bikesharing & 0.5 \euro & 0 \euro & 0.01 \euro \\ \hline
Carsharing & 13 \euro & 0.1 \euro & 0.01 \euro \\ \hline
e-scootersharing & 2,5 \euro & 0.1 \euro & 0.01 \euro \\ \hline
\end{tabular}
\caption{Grid network: mobility services' costs}
\label{tab:costs}
\end{table}



As argued in Section \ref{cha:idea}, understanding how the proposed SMIs impact both commuters and the transport system requires a comparative experiment. In particular, considering that commuters in this scenario are used to move by private cars, first we simulate the current urban mobility (i.e. the \textbf{pre-SMI} simulation), then we simulate each SMI separately (i.e. \textbf{SMI-1}, \textbf{SMI-2} and \textbf{SMI-3} simulations). Afterwards, we compare these simulations with respect to the following variables: traveled distance, travel time, $CO_2$ emissions, cost of mobility and urban traffic levels; all these values are collected during the last iteration of each simulation. The first 4 variables are intended as per-capita indicators (averaged values) and summarize the traveling experience of commuters, whereas the last one can be seen as a performance measure of the urban system.


\subsection{Experimental results}\label{ssec:gridresults}

In this section, we show and discuss the results obtained from our simulations to compare them according to the just mentioned indicators. We also provide some interesting insights concerning the impact of the 3 SMIs on people acceptance and on mobility resource usage levels.  


\begin{figure*}[ht]
    \minipage{0.33\textwidth}
        \includegraphics[width=\linewidth]{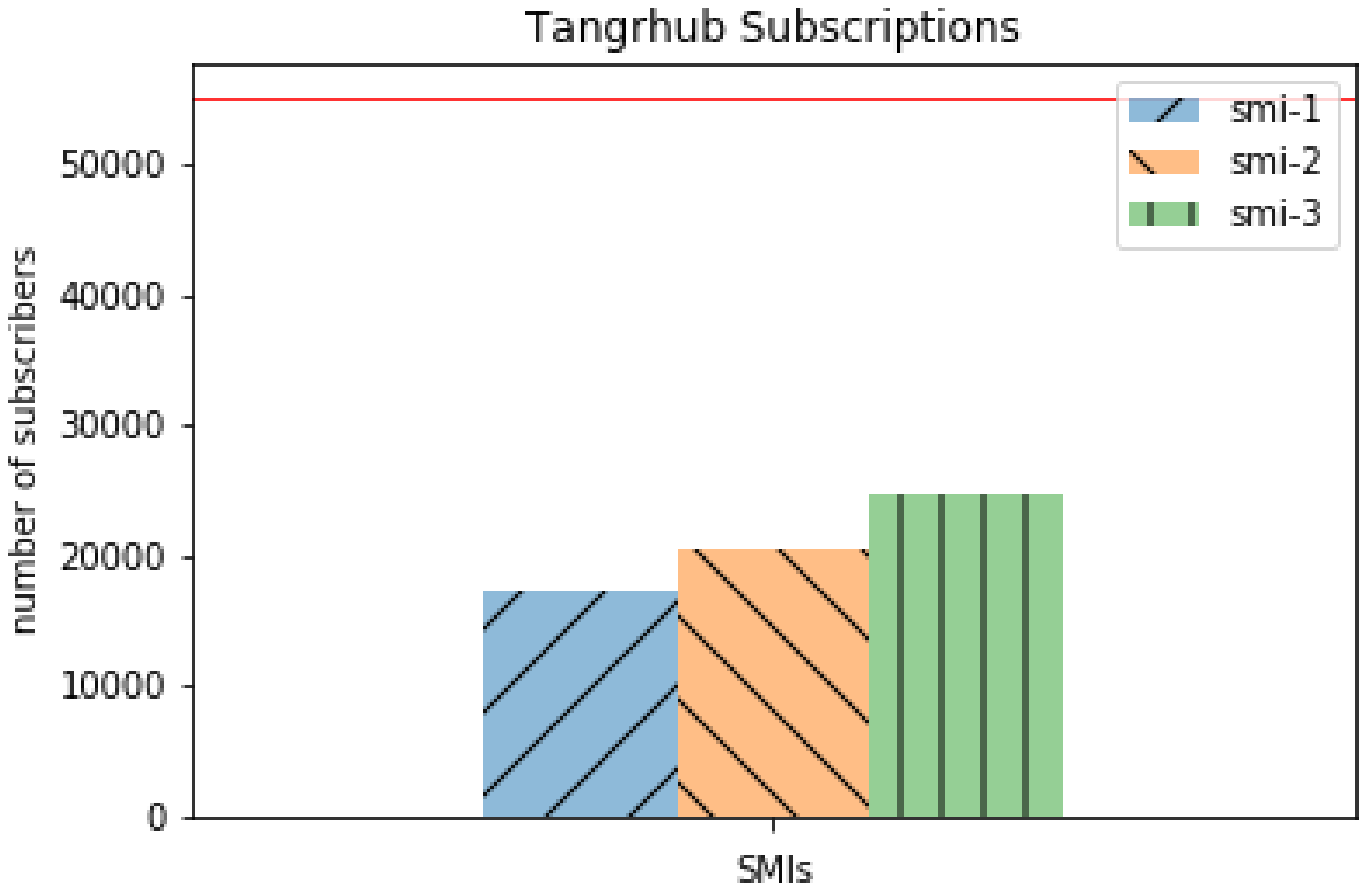}
       \vspace{-1.5em}
        \caption{Tangrhubs subscriptions}
        \label{fig:subscriptions}
    \endminipage\hfill
    \minipage{0.33\textwidth}
        \includegraphics[width=\linewidth]{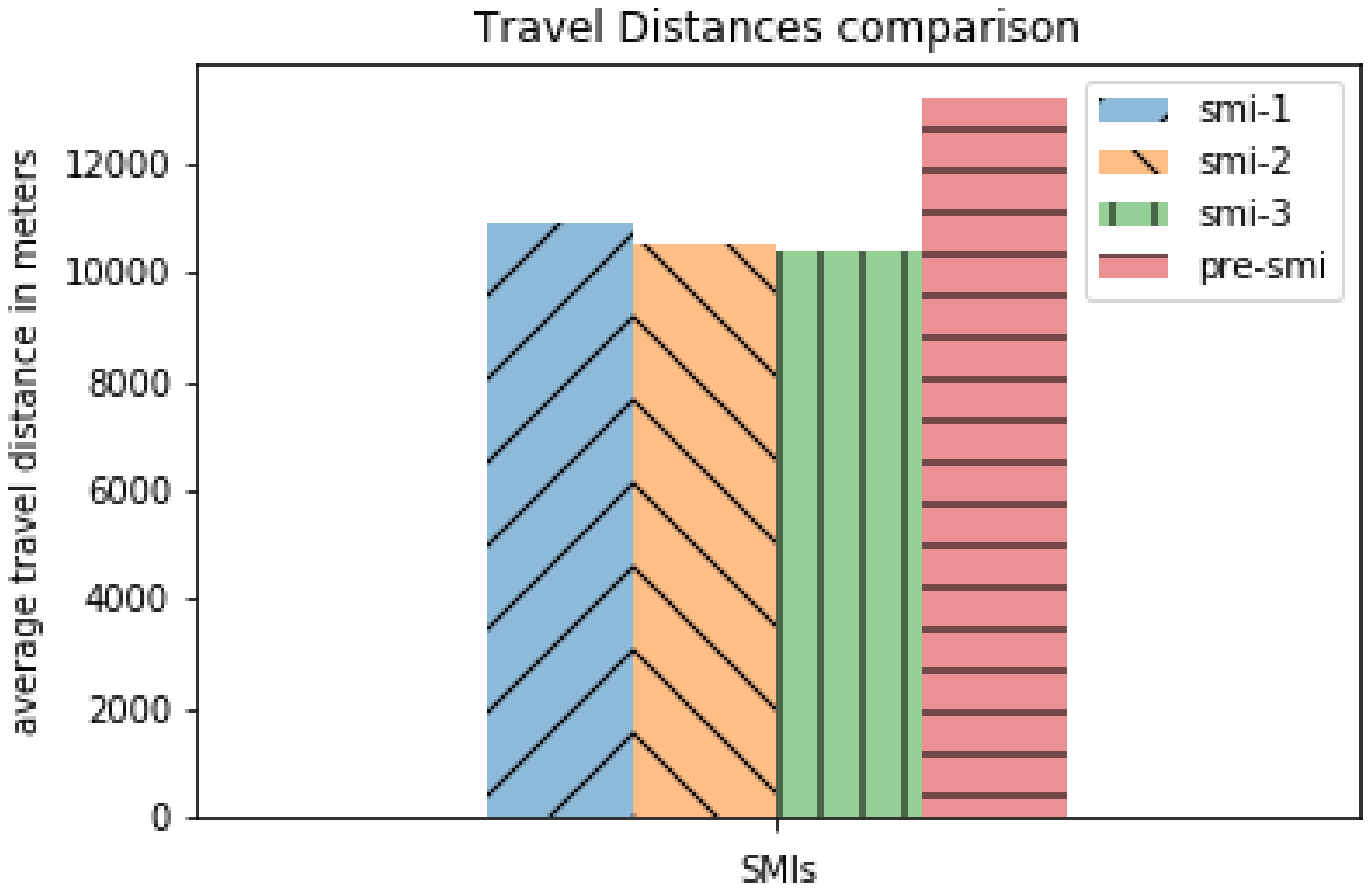}
        \vspace{-1.5em}
        \caption{Traveled distances comparison}
        \label{fig:distancesPlay}
    \endminipage\hfill
    \minipage{0.33\textwidth}%
        \includegraphics[width=\linewidth]{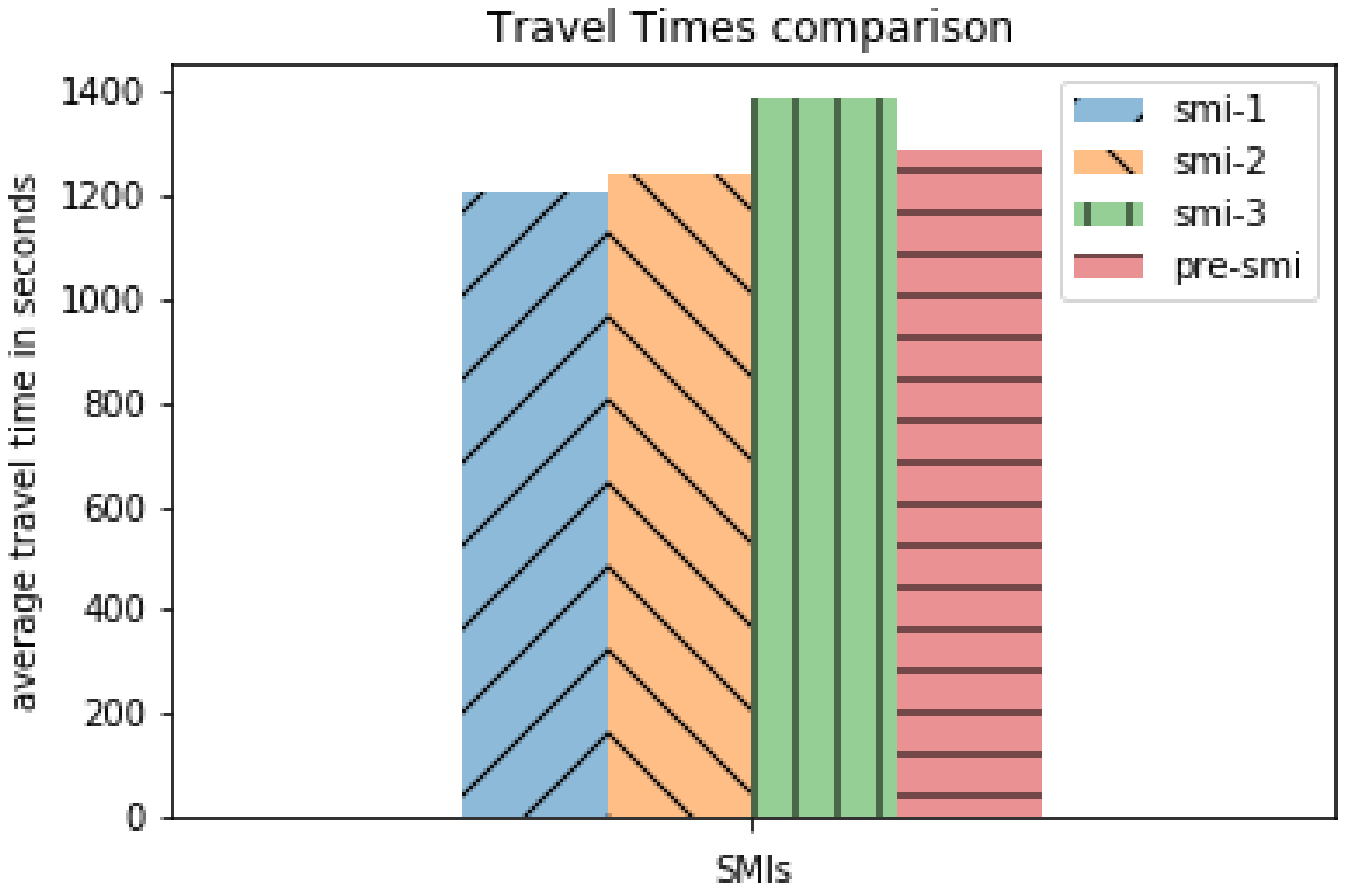}
        \vspace{-1.5em}
        \caption{Travel times comparison}
        \label{fig:timesPlay}
    \endminipage
\end{figure*}

\begin{figure*}[ht]
    \minipage{0.33\textwidth}
        \includegraphics[width=\linewidth]{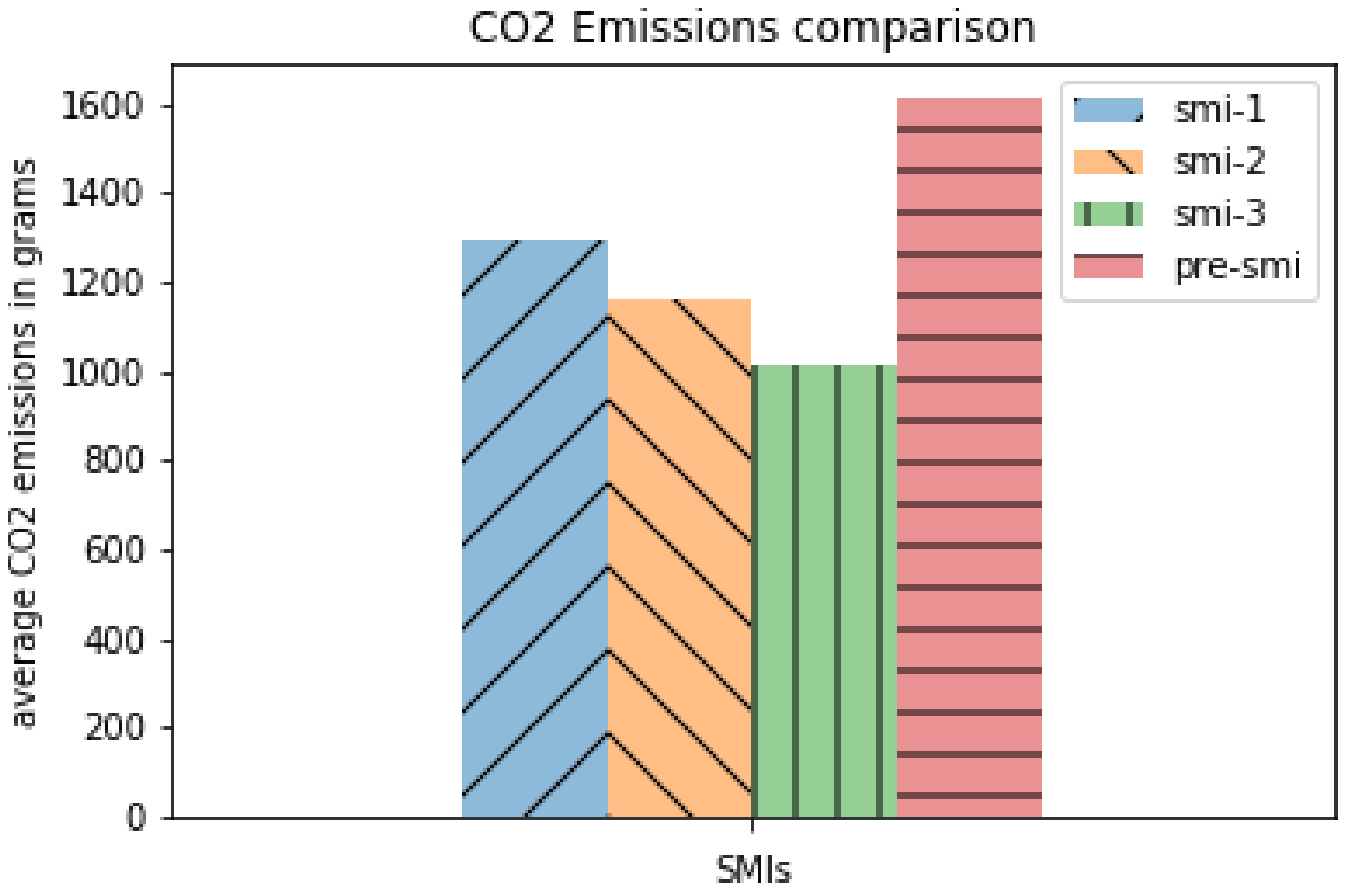}
        \vspace{-1.5em}
        \caption{Emissions comparison}
        \label{fig:emissionsPlay}
    \endminipage\hfill
    \minipage{0.33\textwidth}
        \includegraphics[width=\linewidth]{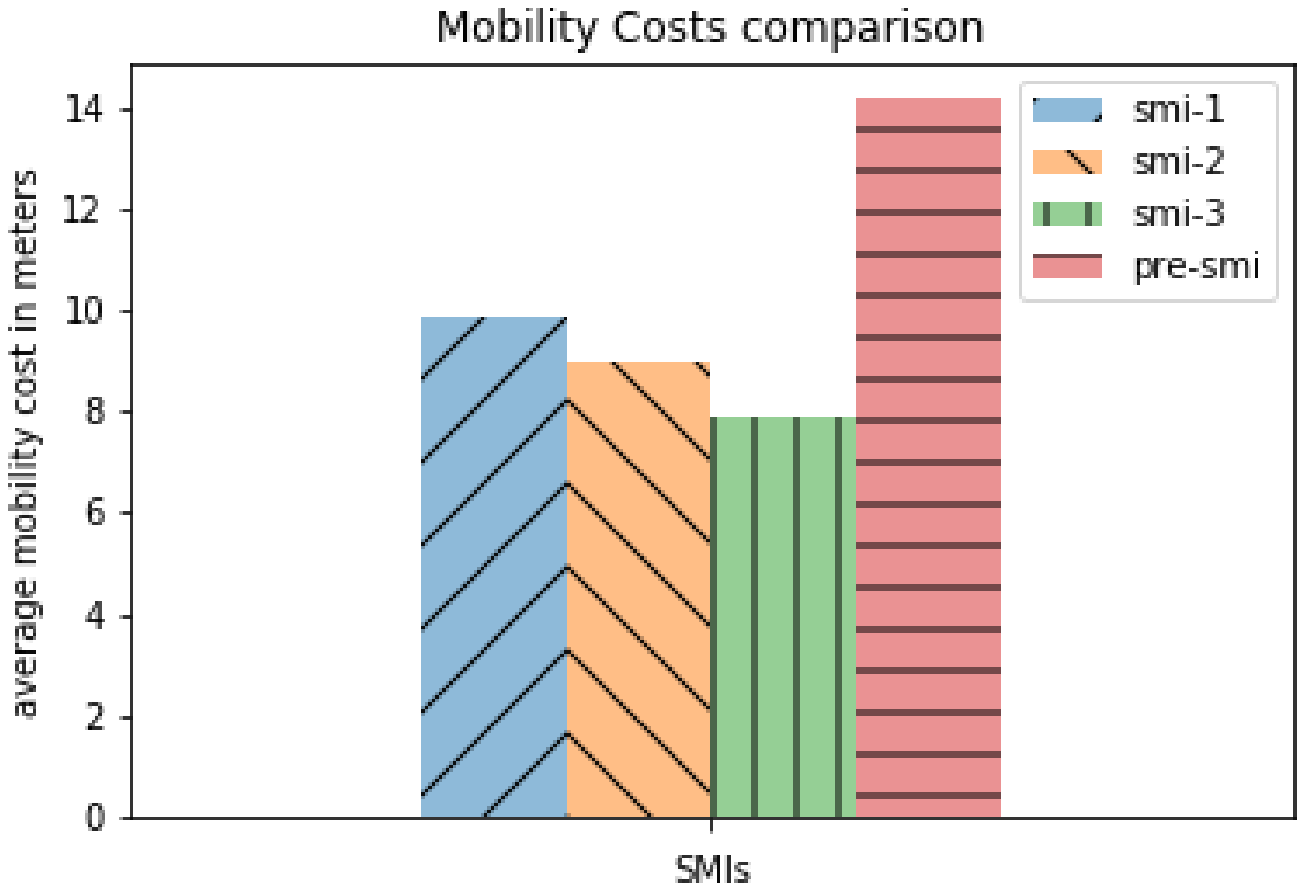}
        \vspace{-1.5em}
        \caption{Mobility costs comparison}
         \label{fig:costsPlay}
    \endminipage\hfill
    \minipage{0.33\textwidth}%
        \includegraphics[width=\linewidth]{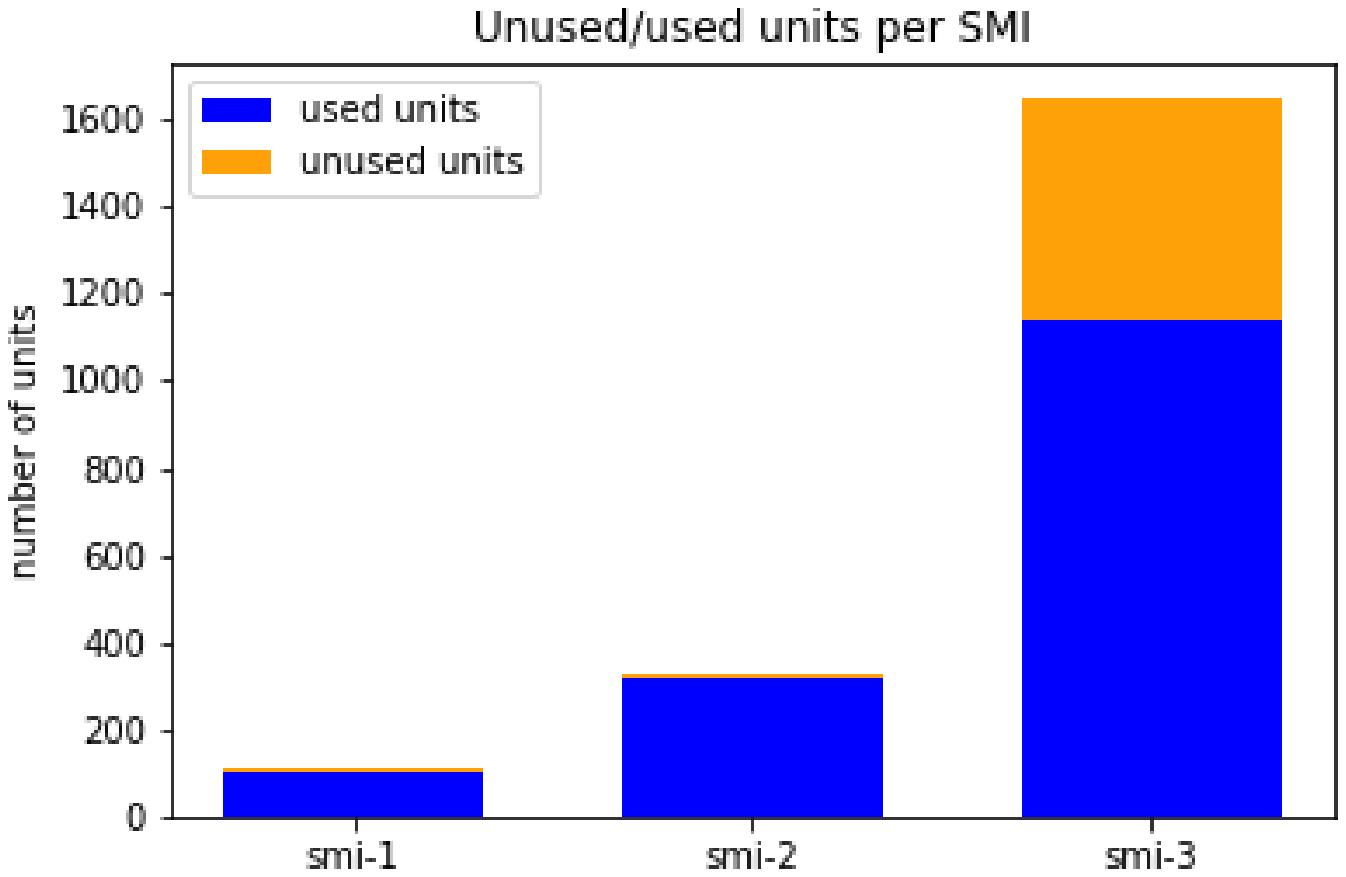} 
        \vspace{-1.5em}
        \caption{Mobility resources usage}
        \label{fig:resourcesUsage}
    \endminipage
\end{figure*}

\subsubsection{Number of \smarthubs' subscribers}
A \emph{subscriber} is a person who uses the mobility services provided by \smarthubs. This value may measure the success of a SMI in terms of people acceptance.
As noticeable in Figure \ref{fig:subscriptions}, the number of subscribers increase as the SMI has more mobility resources, where the horizontal line denotes the entire population.

\subsubsection{Commuters' performance measures}
The first variable involved in the comparison is the \textit{traveled distance} of commuters shown in Figure \ref{fig:distancesPlay}. In the 3 SMI simulations, commuters are expected to travel shorter distances than the pre-SMI ones even if the differences are not so marked.
For what concerns \textit{travel times} (Figure \ref{fig:timesPlay}), in the simulations of SMI-1 and SMI-2 commuters spend less time traveling. This is a good indicator of the effectiveness of the SMIs. Conversely, in SMI-3 commuters spend much more time traveling than before. This indicates that SMI-3 has some problems either in the configuration or in moving people respect to the pre-SMI.
%

%
%
%
Concerning the comparison of \textit{$CO_2$ emissions} produced by commuters, we found that the carbon footprint of the 3 SMIs simulations tends to decrease in directly proportional way with the number of subscribers. Therefore, we can affirm that the more the SMI satisfies a large section of the population, the more the simulation becomes eco-friendly if we use green vehicles.
SMI-1, SMI-2 and SMI-3 reduce ${CO}_2$ respectively of 20\%, 25\% and 35\%.
Besides the environmental impact, we also found that there exists an inverse relationship between the number of subscriptions and the daily \textit{costs of mobility}. As can be seen in Figure \ref{fig:costsPlay}, a commuter in the pre-SMI simulation spends on average \euro{13.5} a day for traveling, whereas a commuter can satisfy his/her needs with a lower expense of \euro{10}, \euro{9} and \euro{8} in the SMI-1, SMI-2 and SMI-3 respectively.






\subsubsection{Mobility resources usage} 

\begin{figure}[ht]
    \centering
    \includegraphics[width=.75\columnwidth]{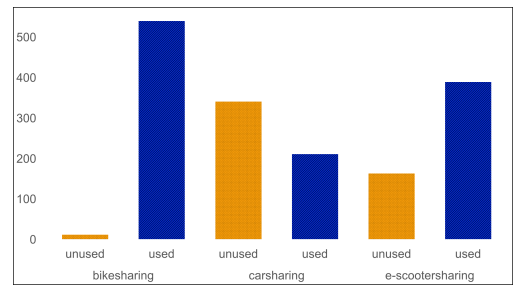}
    \caption{Mobility resource usage SMI-3}
    \label{resource-split-smi3}
\end{figure}

\simu can provide stats concerning the level of mobility resources usage of the SMIs (Figure \ref{fig:resourcesUsage}). The analysis of these data allows to understand if a SMI is efficiently configured, so as to refine it for obtaining similar results with fewer mobility resources. In our case, it turns out that SMI-1 and SMI-2 are properly configured and they resources are used. SMI-3 have a large number of unused vehicles,  we can reduce its fleets in other simulation attempt. A closer look to the resource usage of SMI-3 in Figure \ref{resource-split-smi3} shows the incorrect sizing for the car and scooter services highlighting however the right usage of bikes.

Gathering together all the results, we can conclude that a properly configured SMI helps reducing several urban problems, like traffic congestion levels and consequently air pollution. 
The experiment conclude that SMI-2 shortly reduced distances by 20\% mantaining substantially the same travel times but lowering significantly emissions and costs. Moreover its application actually use all the resources associated with services. 
The same conclusion can be made for SMI-1 also if the benefit is less noticeable.
SMI-1 and SMI-2 could be evaluated in relation to their implementation costs by a urban planner and a decision maker.
SMI-3 on the contrary increases the travel time while maintaining important benefits in travel distances, emission and costs. However, its implementation requires many  resources, many of which are left unused.

The 3 simulations of this experiment took about two and a half hours each, with $110$ iterations on a Linux machine with an i7-7700k CPU @ 4.8GHz and 16GB RAM. We used in the test the whole 56000 agents population. Running a mid-sized scenario with a real population of 500K inhabitants should be done scaling the population to 10\% as suggested by \cite{matsimbook} thus such data makes us confident on feasibility of \simu simulations also for scenarios larger.


\section{Related Work}
\label{cha:related}

To the best of our knowledge, the state of the art does not provide any easy-to-use tool for assessing the impacts of smart mobility initiatives, in particular when several mobility services are considered.
This section thus aims at presenting and discussing the main relevant studies sharing our intent. 



\revisedtext{We present simulation-based studies that} can be defined as in-silico experiments in which scientists investigate how the introduction of a certain smart mobility service could improve the urban transport system of a chosen area of interest. These studies are performed by extending traffic simulators, like MATSim \cite{matsimbook} and SUMO \cite{behrisch2011sumo}, and the rigorous nature of such experiments can guarantee the reliability of their results. 
For instance, in \cite{bischoff2016autonomous}, Bischoff et al. extended MATSim in order to investigate a city-wide replacement of private cars with variously sized Autonomous Taxicab (AT) fleets. Results showed that a fleet of 100,000 AT vehicles could satisfy all the inner city trip demand with: an average waiting time for a vehicle of under three minutes at most times of the day, and under five minutes during peak hours.
%
In \cite{Balac2014}, Balac et al. simulated the introduction of two different carsharing initiatives: round-based and one-way. The first one allows users to pick-up a car from a nearby station and return it later to the same location. The second one is similar to the round-based type, but it allows one to park the car at the closest station from the destination. The simulation was performed by means of a MATSim extension, limiting the scenario to the centre of Zurich. Results proved that the round-based carsharing is mostly chosen for shopping and leisure trips, whilst one-way for commuting ones.


All these studies bring insightful results with regard to how a transportation system is expected to improve after the introduction of a smart mobility service. This is made possible by the scientific nature of such traffic simulations, as showed by the MATSim team in \cite{matsimbook}. However, each of these studies deals with a single solution, so there still is no common framework to assess the impacts and performance of a range of heterogeneous smart mobility services \cite{Zavitsas2011} as well as an holistic and interrelated vision of these actions \cite{McGrath2016}. Furthermore, each study pertains to a specific geographical area (e.g. Berlin for AT, Zurich for carsharing) and the technical organization of the corresponding extension is not flexible enough to be adapted to other areas. Finally, it is worth mentioning that only IT-skilled users would be able to customise such extensions because of their in-code nature.

Other than MATSim and SUMO, traffic simulators like SimMobility \cite{adnan2016simmobility}, Vissim \cite{Fellendorf2010} and SMART (Scalable Microscopic Adaptive Road Traffic Simulator) \cite{RamamohanaraoSMART} do not currently provide relevant studies on the impacts of smart mobility solutions.

\section{Conclusions and Future Work}
\label{cha:conclusions}

Understanding how urban mobility is expected to evolve after the introduction of new smart mobility services is a crucial task in the Urban Planning field. Local public authorities and mobility service providers currently design mobility initiatives according to common heuristics and best practices; these approaches cannot be expected to generalize to every geographical context due to the complexity and the diversity of urban systems. Thus, deploying a mobility initiative is the only way to get measure ex post, but this also comes with potential risks since a failing initiative would result in a considerable waste of resources and trust.


To address this problem we introduced \simu, an agent-based simulation framework that allows users to assess impacts and performances of a mobility initiative within an urban area of interest. \simu performs comparative experiments between, before, and after the introduction of a mobility initiative, approximating real-world urban dynamics by adopting reinforcement learning techniques. The computational nature of these experiments makes it easy to support urban mobility decisions permitting to reduce costs and risks.

\simu is still under active development and improvement, nonetheless the current version already permitted to run meaningful experiments that provided positive results on the usefulness and the potentialities of the simulator.
In particular, users can measure the impacts of a simulated smart mobility initiative with respect to: urban traffic levels, ${CO}_2$ emissions, traveled distances, travel times, land use levels, cost of mobility, number of adopters and resources usage level. Thus, it is up to the user evaluate which variables are more relevant for understanding whether or not an initiative is in line with his objectives.
\revisedtext{%
The experiment we shown help urban planner to consider future initiatives and policies. SMI-1 is the cost effective solution impacting significantly $CO_2$ emissions and personal costs. SMI-2 is the most powerful initiative able to lower again that values while offering a variety of services to the commuters. SMI-3 is clearly oversized and the improvements made possible by its use are not justified by implementation costs and resource unused rate. From these, planners could refine SMI-1 and SMI-2 to arrive at a simulated city planning useful to the decision makers.}

Planned future work includes the extension of the current scoring function with additional traveling comfort criteria  to measure the comfort of a traveling experience with a certain vehicle to let the commuter agents evaluate a mobility service as a whole. We are also  working on additional evaluations, taking into account more complex scenarios.





\ifCLASSOPTIONcaptionsoff
  \newpage
\fi



\bibliographystyle{IEEEtran}
\bibliographystyle{authordate1}
\bibliography{main.bib}
\end{document}